\documentstyle[aps,12pt,manuscript]{revtex}
\begin{document}
\preprint{INJE-TP-97-3, hep-th/9708099}
\def\overlay#1#2{\setbox0=\hbox{#1}\setbox1=\hbox to \wd0{\hss #2\hss}#1%
\hskip -2\wd0\copy1}

\title{The role of fixed scalars in scattering off a 5D black hole}

\author{ H.W. Lee$^1$, Y.S. Myung$^1$ and  Jin Young Kim$^2$ }
\address{$^1$Department of Physics, Inje University, Kimhae 621-749, Korea\\
$^2$ Department of Physics, Kunsan National University, Kunsan 573-701, Korea}

\maketitle
\vskip 1.5in

\begin{abstract}
We discuss the role of fixed scalars($\nu,\lambda$) in 
scattering off a five-dimensional balck hole.  The issue is to 
explain the disagreement of the greybody factor for $\lambda$ 
between the semiclassical and effective string calculations.  
In the effective string approach, this is related to the 
operators with dimension (3,1) and (1,3).  On the semiclassical 
calculation, this originates from a complicated mixing 
between $\lambda$ and other fields.  Hence it may depend on the 
decoupling procedure.  
It is shown that $\lambda$ depends on the gauge choices such as the harmonic, 
dilaton gauges, and the Krasnitz-Klebanov setting for $h_{\mu\nu}$.  
It turns out that $\nu$ plays a role of test field well, 
while the role of $\lambda$ is obscure.

\end{abstract}

\newpage
\section{Introduction}
\label{introduction}
Recently there has been a great progress in a certain class of 
five-dimensional(5D) black holes 
with three U(1) charges.  This progress was achieved in both the 
Bekenstein-Hawking entropy($S_{BH}$) and absorption 
cross section($\sigma_{abs}$).  
The semiclassical calculations of cross section(greybody factor) 
in the extremal and near extremal black holes are 
important to compare them with the result of D-branes.

Apart from counting the microstates\cite{Vaf} of black hole 
through the D-brane physics, a dynamical consideration becomes an  
important issue\cite{Mal,Das,Cal,Kra97,Das2}.
This is so because the greybody factor for the black hole arises as a 
consequence of the gravitational potential barrier surrounding the 
horizon.  That is, this is an effect of spacetime curvature.  In the stringy 
description, their origin comes from the thermal distribution for 
excitations of the D1-D5 bound state.  Together with the 
Bekenstein-Hawking entropy, this seems to be a strong hint of a 
deep and mysterious connection between curvature and statistical 
mechanics\cite{Emp}.  
The cross section calculation for a minimally coupled scalar 
is straightforward in both semiclassical and effective string models.  
The s-wave cross section is not sensitive 
to the moduli and energy($\omega$). 
 This depends on the area of horizon\cite{Das}. 
However, this is true when the area of the horizon is not zero,
e.g., for a 5D black hole with three charges.  When a 5D black hole 
has only two charges, the absorption cross section 
depends on both moduli and energy\cite{Emp}. 

A better test of the agreement between semiclassical and effective string 
calculations is provided by the fixed scalars.  The effective string 
calculation is well performed in the 
dilute gas approximation.  But the semiclassical calculations are 
difficult even for the dilute gas limit, 
because of a complicated mixing between fixed 
scalars and other fields (metric and gauge fields).
One of fixed scalars($\nu$) is coupled to an operator of 
dimension (2,2) in the effective string model.  
When D1-brane charge($Q_1$) is equal to D5-brane charge($Q_5$), 
the string calculation of 
$\sigma_{abs}$ yields the precise agreement with the 
semiclassical greybody factor\cite{Cal}.  
But the greybody factor for the 
other ($\lambda$) is not in agreement for $Q_1=Q_5$.  
This disagreement is related to the presence of the 
chiral operators with (3,1) and (1,3) in the effective string approach.  
On the other hand, this originates from a complicated mixing between 
$\lambda$ and other fields in the semiclassical calculation.  
Thus it may depend on the decoupling procedure.  Here we deal 
mainly with this problem.  

In this paper, we shall perform a complete, semiclassical analysis 
for a 5D black hole with three U(1) charges.  
This is similar to the 4D N=4 black hole with two 
U(1) charges\cite{Kal},
which provides us a simple model for getting
the s-wave cross section of the fixed scalar\cite{Kol}.
Here we consider all perturbing equations around a 5D black hole
to find the consistent solution.  
In the s-wave calculation two fixed scalars are 
physically propagating modes, whereas other fields become the redundant ones.  
Hence our main task is to decouple the fixed scalars from all other fields.
In order to achieve this, we first consider the general perturbation for 
the graviton $h_{\mu\nu}$.  We choose either the 
harmonic gauge($\nabla_\mu {\hat h}^{\mu\nu}=0, {\hat h}^{\mu\nu} = 
h^{\mu\nu} - {1 \over 2} g^{\mu\nu} h$) or the dilaton gauge
($\nabla_\mu {\hat h}^{\mu\rho}=h^{\mu\nu} \Gamma^\rho_{\mu\nu}$). 
It turns out that for $Q_1=Q_5$, $\nu$ is independent 
of the gauge-fixing, while $\lambda$ depends on the gauge choice. 
This may explain the agreement of greybody factor for $\nu$ and 
disagreement for $\lambda$.  For an explicit calculation we choose 
to the Krasnitz-Klebanov(K-K) setting for $h_{\mu\nu}$ 
as in Ref.\cite{Kra97}.  
This is not suitable for studying 
the higher angular momentum modes ($l \ge 1$)\cite{Mal1}.  
In order to study higher modes, we need to consider 
the general perturbation as in Ref.\cite{Hol,Cha83}.

The organization of our paper is as follows.  In Sec. \ref{5dblackhole}, 
we review  the revelant part of a 
5D black hole briefly.  We set up the perturbation for all fields around a 
5D black hole solutions in Sec. \ref{perturbation}.  
The s-wave absorption cross section is calculated in 
Sec. \ref{propagation}.  Finally, we discuss 
the role of fixed scalars as the test fields 
in Sec. \ref{discussion}.

\section{5D Black Holes}
\label{5dblackhole}
Here we consider a class of 5D black holes representing 
the bound state of $n_1(={VQ_1 \over g})$ D1 strings and 
$n_5(={Q_5 \over g})$ D5-branes compactified on 
a $T^5(=T^4 \times S^1$).  This black hole can also be obtained as a solution 
to the semiclassical action of type IIB superstring 
compactified on $T^5$.
The effective action for a 5D black hole with three 
charges is given by\cite{Cal,Kra97}
\begin{equation}
S= {1 \over {2 \kappa_5^2}} \int d^5 x \sqrt{-g} \left \{ R - 
{4 \over 3} (\nabla \lambda)^2 - 4 (\nabla \nu)^2 
-{1 \over 4} e^{{8 \over 3} \lambda} F^{(K)2} 
-{1 \over 4} e^{-{4 \over 3} \lambda + 4 \nu} F^2
-{1 \over 4} e^{-{4 \over 3} \lambda -4 \nu} H^2 \right \} ,
\label{eq-action}
\end{equation}
where $F^{(K)}_{\mu\nu}$ is the Kaluza-Klein(KK) field strength along the 
string direction($S^1$), $F_{\mu\nu}$ is the electric components of 
the Ramond-Ramond(RR) two-form and $H_{\mu\nu}$ is dual to 
the magnetic components of the RR 
two-form.  Here we omit the analysis of the 6D dilaton $\phi_6$, 
since it is just a minimally 
decoupled scalar.  On the other hand, the scalars $\nu$ and $\lambda$ 
interact with the gauge fields and are examples of the fixed scalar.  $\nu$ is
related to the scale of the internal torus($T^4$), 
while $\lambda$ is related to 
the scale of the KK circle($S^1$).  $\kappa_5^2$ is the 5D  gravitational 
coupling constant ($\kappa_5^2=8 \pi G_N^5, G_N^5$=5D Newtonian 
constant).  This can be determined by $G_N^5 = {G_N^{10} \over V_5} =
{8 \pi^6 g^2 \over (2\pi)^5 VR} = {\pi g^2 \over 4 VR}$ 
with $V=R_5R_6R_7R_8$(volume of $T^4$), 
$R=R_9$(radius of $S^1$), $\alpha'=1,$ and 
$g$(=10D string coupling constant).  
We wish to follow the MTW conventions\cite{Mtw}.

The equations for action (\ref{eq-action}) is given by
\begin{eqnarray}
&&R_{\mu \nu} - {4 \over 3} \partial_\mu \lambda \partial_\nu \lambda 
- 4 \partial_\mu \partial_\nu \nu 
- e^{{8 \over 3} \lambda} \left ( {1 \over 2} F^{(K)}_{\mu \rho}
  F^{(K)~\rho}_{~~~~\nu} - {1 \over 12} F^{(K)2} g_{\mu \nu} \right )
\nonumber \\
&&~~~~- e^{-{4 \over 3} \lambda + 4 \nu} \left ( 
{1 \over 2}F_{\mu \rho} F^{~\rho}_\nu - {1 \over 12} F^2 g_{\mu \nu} \right ) 
- e^{-{4 \over 3}\lambda -4 \nu} \left ( {1 \over 2} H_{\mu \rho} 
  H^{~\rho}_\nu - {1 \over 12} H^2 g_{\mu \nu} \right ) = 0,
\label{eq-ricci} \\
&& 8 \nabla^2 \nu - e^{-{4 \over 3} \lambda+4 \nu} F^2
+ e^{-{4 \over 3}\lambda -4 \nu} H^2 = 0 \label{eq-nu}\\
&& 8  \nabla^2 \lambda - 2 e^{{8 \over 3} \lambda} F^{(K)2}
+ e^{-{4 \over 3} \lambda +4 \nu} F^2 
+ e^{-{4 \over 3} \lambda -4 \nu} H^2 = 0, \label{eq-lambda}\\
&&\nabla_\mu \left ( e^{{8 \over 3} \lambda} F^{(K)\mu \nu} \right ) =0, 
   \label{eq-fk} \\
&&\nabla_\mu \left ( e^{-{4 \over 3} \lambda+4\nu} F^{\mu \nu} \right ) =0, 
   \label{eq-f} \\
&&\nabla_\mu \left ( e^{-{4 \over 3} \lambda-4\nu} H^{\mu \nu} \right ) =0. 
   \label{eq-h} 
\end{eqnarray}
In addition, we need the remaining Maxwell equations as three 
Bianchi identities\cite{Hol,Gub} 
\begin{equation}
\partial_{[\mu}F^{(K)}_{~~~~\rho\sigma]}=
\partial_{[\mu}F_{\rho\sigma]}=\partial_{[\mu}H_{\rho\sigma]}=0.
\label{bianchi}
\end{equation}

The black hole solution is given by the background metric  
\begin{equation}
ds^2 = - d f^{-{2 \over 3}}dt^2 +d^{-1}f^{1 \over 3} dr^2 + 
r^2 f^{1 \over 3}d \Omega^2_3
\label{defg}
\end{equation}
and 
\begin{eqnarray}
&&e^{2\bar \lambda} = { f_K \over \sqrt{f_1 f_5}},~~~~
e^{4\bar \nu} = { f_1 \over f_5}, ~~~~f=f_1 f_5 f_K, \label{sol-scalar} \\
&&\bar F^{(K)}_{t r} = { 2 Q_K \over {r^3 f_K^2}}, ~~~~
\bar F_{t r} = { 2 Q_1 \over {r^3 f_1^2}}, ~~~~
\bar H_{t r} = { 2 Q_5 \over {r^3 f_5^2}}. \label{sol-vector}
\end{eqnarray}
Here four harmonic functions are defined by
\begin{equation}
f_1= 1 +{ r_1^2 \over r^2}, ~~~~ f_5= 1 +{ r_5^2 \over r^2},~~~~
f_K = 1 + { r_K^2 \over r^2},~~~~ d = 1 - { r_0^2 \over r^2},
\label{harmonics}
\end{equation}
with $r_i^2=r_0^2 \sinh^2\sigma_i,~ i=1,5,K$.
$Q_1$, $Q_5$ and $Q_K$ are related to 
the characteristic radii $r_1$, $r_5$, 
$r_K$ and the radius of horizon $r_0$ as
\begin{eqnarray}
&&Q_i = {1 \over 2} r_0^2 \sinh 2 \sigma_i,~ Q_i^2 = 
r_i^2 (r_i^2 + r_0^2), ~r_i^2 = \sqrt{Q_i^2 + {r_0^4 \over 4}} - 
{r_0^2 \over 2}. \label{eq-charges}
\end{eqnarray}

The background metric (\ref{defg}) is just a 5D Schwarzschild one 
with time and space components rescaled by different powers of $f$.  The event 
horizon (outer horizon) is clearly at $r=r_0$.  When all three charges are 
nonzero, the surface of $r=0$ becomes a smooth inner horizon (Cauchy horizon).  
If one of the charges is zero, the surface of $r=0$ becomes singular.  The 
extremal case corresponds  to the limit of $r_0 \rightarrow 0$ with the boost 
parameters $\sigma_i \rightarrow \pm \infty$ keeping 
$Q_i$ fixed.  Here one has $Q_1 = r_1^2,~Q_5 = r_5^2$, 
and $Q_K=r_K^2$.  
In this work we are very interested in the limit 
of $r_0, r_K \ll r_1, r_5$, which is called 
the dilute gas approximation.  This corresponds to the near-extremal black hole 
and its thermodynamic quantities are given by
\begin{eqnarray}
&&M_{next} = {2 \pi^2 \over \kappa_5^2} \left [ r_1^2 +
r_5^2 + {1 \over 2}r_0^2\cosh 2 \sigma_K \right ],\label{next-mass}\\
&&S_{next} = { 4 \pi^3 r_0 \over \kappa_5^2} r_1 r_5\cosh \sigma_K
, \label{next-entropy}\\
&&{1 \over T_{H,next}} = {2 \pi \over r_0} r_1 r_5 \cosh \sigma_K.
\label{next-hawking}
\end{eqnarray}
The above energy and entropy are actually those of a gas of 
massless 1D particles.  
In this case the effective temperatures of the left and right moving 
string modes are given by
\begin{equation}
T_L = {1 \over 2 \pi} \left ( {r_0 \over r_1 r_5} \right ) e^{\sigma_K},~~
T_R = {1 \over 2 \pi} \left ( {r_0 \over r_1 r_5} \right ) e^{-\sigma_K}.
\label{eff-temp}
\end{equation}
The Hawking temperature is given by their harmonic average
\begin{equation}
{2 \over T_H} = {1 \over T_L} + {1 \over T_R}.
\label{eff-hawking}
\end{equation}

\section{Perturbation Analysis}
\label{perturbation}
Here we start with the perturbation 
around the black hole background as\cite{Lee}
\begin{eqnarray}
&&F^{(K)}_{tr} = \bar F^{(K)}_{tr} + {\cal F}^{(K)}_{tr} = 
\bar F^{(K)}_{tr} [1 + {\cal F}^{(K)}(t,r,\chi, \theta,\phi)],
\label{ptrFk} \\
&&F_{tr} = \bar F_{tr} + {\cal F}_{tr} = 
\bar F_{tr} [1 + {\cal F}(t,r,\chi,\theta,\phi)],
\label{ptrF} \\
&&H_{tr} = \bar H_{tr} + {\cal H}_{tr} = 
\bar H_{tr} [1 + {\cal H}(t,r,\chi,\theta,\phi)],
\label{ptrH} \\    
&&\lambda = \bar \lambda + \delta\lambda(t,r,\chi,\theta,\phi),  
\label{ptr-lambda}  \\  
&&\nu = \bar \nu + \delta\nu(t,r,\chi,\theta,\phi),  
\label{ptr-nu}  \\  
&&g_{\mu\nu} = \bar g_{\mu\nu} + h_{\mu\nu}.
\label{ptrg}
\end{eqnarray}
Here $h_{\mu\nu}$ is given by
\begin{equation}
h^\mu_{~\nu} = 
\left [
\begin{array}{ccccc}
h_1& h_3 & 0 & 0 & 0 \\
-d^2 h_3 /f & h_2 & 0 & 0 & 0 \\
0 & 0 & h^\chi_{~\chi} & h^\chi_{~\theta} & h^\chi_{~\phi} \\
0 & 0 & h^\theta_{~\chi} & h^\theta_{~\theta} & h^\theta_{~\phi} \\
0 & 0 & h^\phi_{~\chi} & h^\phi_{~\theta} & h^\phi_{~\phi} 
\end{array}
\right ]
\label{ptr-h}
\end{equation}
This seems to be general for the s-wave calculation.

One has to linearize (\ref{eq-ricci})-(\ref{eq-h}) in order to obtain 
the equations governing the perturbations as
\begin{eqnarray}
&&  \delta R_{\mu\nu} (h)  
-{4 \over 3}(\partial_\mu \bar \lambda \partial_\nu \delta \lambda +
\partial_\mu \delta \lambda \partial_\nu \bar \lambda)
-4(\partial_\mu \bar \nu \partial_\nu \delta \nu +
\partial_\mu \delta \nu \partial_\nu \bar \nu) \nonumber \\
&&~~~~+{1 \over 2} e^{{8 \over 3} \bar \lambda} 
\bar F^{(K)}_{\mu \rho} \bar F^{(K)}_{\nu \alpha}h^{ \rho\alpha}   
-e^{{8 \over 3} \bar \lambda}\bar F^{(K)}_{\mu \rho} {\cal F}^{(K)~\rho}_{~~~~\nu} 
-{4 \over 3}e^{{8 \over 3} \bar \lambda}
\bar F^{(K)}_{\mu \rho} \bar F^{(K)~\rho}_{~~~~\nu} \delta\lambda 
\nonumber  \\
&&~~~~+{1 \over 6} e^{{8 \over 3} \bar \lambda} 
\bar F^{(K)}_{\rho\sigma} {\cal F}^{(K)\rho\sigma}\bar g_{\mu\nu}   
-{1 \over 6}e^{{8 \over 3} \bar \lambda}\bar F^{(K)}_{\rho\kappa} 
\bar F^{(K)~\kappa}_{~~~~\eta} h^{\rho\eta} \bar g_{\mu\nu}
+{2 \over 9}e^{{8 \over 3} \bar \lambda}
\bar F^{(K)2}\bar g_{\mu\nu} \delta\lambda 
+{1 \over 12}e^{{8 \over 3} \bar \lambda} \bar F^{(K)2} h_{\mu\nu} 
\nonumber  \\
&&~~~~+{1 \over 2} e^{-{4 \over 3} \bar \lambda + 4 \bar \nu} 
\bar F_{\mu \rho} \bar F_{\nu \alpha}h^{ \rho\alpha}   
-e^{-{4 \over 3} \bar \lambda + 4 \bar \nu}
\bar F_{\mu \rho} {\cal F}^{~\rho}_{\nu} 
-e^{-{4 \over 3} \bar \lambda+4 \bar \nu}
\bar F_{\mu \rho} \bar F^{~\rho}_{\nu} 
(-{2 \over 3}\delta\lambda +2 \delta \nu )
\nonumber  \\
&&~~~~+{1 \over 6} e^{-{4 \over 3} \bar \lambda+4 \bar \nu} 
\bar F_{\rho\sigma} {\cal F}^{\rho\sigma}\bar g_{\mu\nu}   
-{1 \over 6}e^{-{4 \over 3} \bar \lambda +4 \bar \nu}\bar F_{\rho\kappa} 
\bar F^{~\kappa}_{\eta} h^{\rho\eta} \bar g_{\mu\nu}
\nonumber  \\
&&~~~~+e^{-{4 \over 3} \bar \lambda +4 \bar \nu}
\bar F^2\bar g_{\mu\nu} (-{1 \over 9}\delta\lambda + {1 \over 3} \delta \nu)
+{1 \over 12}e^{-{4 \over 3} \bar \lambda + 4 \bar \nu} \bar F^2 h_{\mu\nu} 
\nonumber  \\
&&~~~~+{1 \over 2} e^{-{4 \over 3} \bar \lambda - 4 \bar \nu} 
\bar H_{\mu \rho} \bar H_{\nu \alpha}h^{ \rho\alpha}   
-e^{-{4 \over 3} \bar \lambda - 4 \bar \nu}
\bar H_{\mu \rho} {\cal H}^{~\rho}_{\nu} 
+e^{-{4 \over 3} \bar \lambda-4 \bar \nu}
\bar H_{\mu \rho} \bar H^{~\rho}_{\nu} 
({2 \over 3}\delta\lambda +2 \delta \nu )
\nonumber  \\
&&~~~~+{1 \over 6} e^{-{4 \over 3} \bar \lambda-4 \bar \nu} 
\bar H_{\rho\sigma} {\cal H}^{\rho\sigma}\bar g_{\mu\nu}   
-{1 \over 6}e^{-{4 \over 3} \bar \lambda -4 \bar \nu}\bar H_{\rho\kappa} 
\bar H^{~\kappa}_{\eta} h^{\rho\eta} \bar g_{\mu\nu}
\nonumber  \\
&&~~~~-e^{-{4 \over 3} \bar \lambda -4 \bar \nu}
\bar H^2\bar g_{\mu\nu} ({1 \over 9}\delta\lambda + {1 \over 3} \delta \nu)
+{1 \over 12}e^{-{4 \over 3} \bar \lambda - 4 \bar \nu} \bar H^2 h_{\mu\nu} 
=0,
\label{linR} \\
&&  \bar \nabla^2 \delta \nu
- h^{\mu\nu} \bar \nabla_\mu \bar \nabla_\nu \bar \nu   
- \bar g^{\mu\nu} \delta \Gamma^\rho_{\mu\nu} (h) \partial_\rho \bar \nu
\nonumber \\
&&~~~~~~-{1 \over 4} e^{-{4 \over 3} \bar \lambda+4 \bar \nu}
\bar F_{\mu \nu} {\cal F}^{\mu \nu}                
+{1 \over 4} e^{-{4 \over 3} \bar \lambda+4 \bar \nu} 
\bar F_{\mu \nu}   \bar F_{\rho}^{~\nu} h^{\mu\rho} 
-{1 \over 8}e^{-{4 \over 3} \bar \lambda+4 \bar \nu} \bar F^2 
( -{4 \over 3}\delta \lambda + 4 \delta \nu)  \nonumber  \\
&&~~~~~~+{1 \over 4} e^{-{4 \over 3} \bar \lambda-4 \bar \nu}
\bar H_{\mu \nu} {\cal H}^{\mu \nu}                
-{1 \over 4} e^{-{4 \over 3} \bar \lambda-4 \bar \nu} 
\bar H_{\mu \nu}   \bar H_{\rho}^{~\nu} h^{\mu\rho} 
-{1 \over 8}e^{-{4 \over 3} \bar \lambda-4 \bar \nu} \bar H^2 
( {4 \over 3}\delta \lambda + 4 \delta \nu)   = 0,
\label{lin-nu}  \\
&&  \bar \nabla^2 \delta \lambda
- h^{\mu\nu} \bar \nabla_\mu \bar \nabla_\nu \bar \lambda   
- \bar g^{\mu\nu} \delta \Gamma^\rho_{\mu\nu} (h) \partial_\rho \bar \lambda
\nonumber \\
&&~~~~~~-{1 \over 2} e^{{8 \over 3} \bar \lambda}
\bar F^{(K)}_{\mu \nu} {\cal F}^{(K)\mu \nu}                
+{1 \over 2} e^{{8 \over 3} \bar \lambda} 
\bar F^{(K)}_{\mu \nu}   \bar F_{~~~~\rho}^{(K)~\nu} h^{\mu\rho} 
-{2 \over 3} e^{{8 \over 3} \bar \lambda} \bar F^{(K)2} 
\delta \lambda  \nonumber  \\
&&~~~~~~+{1 \over 4} e^{-{4 \over 3} \bar \lambda+4 \bar \nu}
\bar F_{\mu \nu} {\cal F}^{\mu \nu}                
-{1 \over 4} e^{-{4 \over 3} \bar \lambda+4 \bar \nu} 
\bar F_{\mu \nu}   \bar F_{\rho}^{~\nu} h^{\mu\rho} 
+e^{-{4 \over 3} \bar \lambda+4 \bar \nu} \bar F^2 
( -{1 \over 6}\delta \lambda + {1 \over 2} \delta \nu)  \nonumber  \\
&&~~~~~~+{1 \over 4} e^{-{4 \over 3} \bar \lambda-4 \bar \nu}
\bar H_{\mu \nu} {\cal H}^{\mu \nu}                
-{1 \over 4} e^{-{4 \over 3} \bar \lambda-4 \bar \nu} 
\bar H_{\mu \nu}   \bar H_{\rho}^{~\nu} h^{\mu\rho} 
-e^{-{4 \over 3} \bar \lambda-4 \bar \nu} \bar H^2 
( {1 \over 6}\delta \lambda + {1 \over 2} \delta \nu)   = 0,
\label{lin-lambda}  \\
&&( \bar \nabla_\mu +{8 \over 3} \partial_\mu \bar \lambda ) 
    ( {\cal F}^{(K)\mu \nu} - \bar F_{~~~~\alpha}^{(K)~\nu} h^{\alpha \mu}
                             - \bar F^{(K)\mu}_{~~~~~\beta} h^{\beta \nu} )
      + \bar F^{(K)\mu \nu} (\delta \Gamma^\sigma_{\sigma \mu} (h)       
      +{8 \over 3} \partial_\mu \delta \lambda) 
 = 0,  \label{linFk} \\
&&(\bar\nabla_\mu -{4 \over 3}\partial_\mu \bar\lambda+4 \partial_\mu \bar \nu) 
    ( {\cal F}^{\mu \nu} - \bar F_{\alpha}^{~\nu} h^{\alpha \mu}
                             - \bar F^{\mu}_{~\beta} h^{\beta \nu} )
      + \bar F^{\mu \nu} (\delta \Gamma^\sigma_{\sigma \mu} (h)       
      -{4 \over 3} \partial_\mu \delta \lambda 
      +4 \partial_\mu \delta \nu) 
 = 0,  \label{linF} \\
&&(\bar\nabla_\mu -{4 \over 3}\partial_\mu \bar\lambda-4 \partial_\mu \bar \nu) 
    ( {\cal H}^{\mu \nu} - \bar H_{\alpha}^{~\nu} h^{\alpha \mu}
                             - \bar H^{\mu}_{~\beta} h^{\beta \nu} )
      + \bar H^{\mu \nu} (\delta \Gamma^\sigma_{\sigma \mu} (h)       
      -{4 \over 3} \partial_\mu \delta \lambda 
      -4 \partial_\mu \delta \nu) 
 = 0,  \label{linH}
\end{eqnarray}
where 
\begin{eqnarray}
&&\delta R_{\mu\nu} (h) = 
 - {1 \over 2} \bar \nabla^2 h_{\mu\nu}   
 -{1 \over 2} \bar \nabla_\mu \bar \nabla_\nu h^\rho_{~\rho}
 + {1 \over 2} \bar \nabla^\rho \bar \nabla_\nu h_{\rho\mu}   
 + {1 \over 2} \bar \nabla^\rho \bar \nabla_\mu h_{\nu\rho},   
\label{delR} \\   
&&\delta \Gamma^\rho_{\mu\nu} (h) = {1 \over 2} \bar g^{\rho\sigma} 
( \bar \nabla_\nu h_{\mu\sigma} + \bar \nabla_\mu h_{\nu\sigma} - \bar \nabla_\sigma h_{\mu\nu} ).
\label{delGam}
\end{eqnarray}
Since we start with full degrees of freedom (\ref{ptr-h}), we 
choose a gauge to study the propagation of fields.
For this purpose $\delta R_{\mu\nu}$ can be transformed into the 
Lichnerowicz operator\cite{Gre94}
\begin{equation}
\delta R_{\mu\nu} = - {1 \over 2} {\bar \nabla}^2 h_{\mu\nu} 
+ {\bar R}_{\sigma ( \nu} h^\sigma_{~\mu)}
- {\bar R}_{\sigma \mu \rho \nu} h^{\sigma \rho} 
+ {\bar \nabla}_{(\nu} {\bar \nabla}_{|\rho|} {\hat h}^\rho_{~\mu)}.
\label{lichnerowicz}
\end{equation}
We have to examine whether there exists any choice of gauge which can simplify 
Eqs.(\ref{lin-nu}) and (\ref{lin-lambda}).
Conventionally we choose the harmonic (transverse) gauge
(${\bar \nabla}_\mu {\hat h}^{\mu\rho} = \bar g^{\mu\nu} 
\delta \Gamma^\rho_{\mu\nu} = 0 $) if one concentrates on 
the propagation of gravitons.

\subsection{Harmonic Gauge}
Considering the harmonic gauge and $Q_1=Q_5$ case, Eqs.(\ref{lin-nu}) and 
(\ref{lin-lambda}) lead to
\begin{eqnarray}
&&{\bar \nabla}^2 \delta \nu + { Q_1^2 \over r^6 f_1^2 f^{1/3}}
  (2 {\cal F} -2 {\cal H} + 8 \delta \nu ) =0, 
\label{nu-harmonic} \\
&&{\bar \nabla}^2 \delta \lambda  
- { d \over f^{1/3} } h^{rr} \partial^2_r \bar \lambda
+ { d \over f^{1/3}} h^{\mu\nu} \Gamma^r_{\mu\nu} \partial_r \bar \lambda
- {2 Q_K^2 \over r^6 f_K^2 f^{1/3}} (h_1 + h_2 - 2 {\cal F}^{(K)} 
          -{8 \over 3} \delta \lambda ) 
\nonumber \\
&&~~~~~~+ { 2 Q_1^2 \over r^6 f_1^2 f^{1/3}}
  ( h_1 +  h_2 -  {\cal F} - {\cal H} + {4 \over 3}\delta \lambda ) =0. 
\label{lambda-harmonic} 
\end{eqnarray}
Now we attempt to disentangle the mixing between ($\delta \nu, \delta \lambda$)
and other fields by using both the harmonic gauge and U(1) 
field equations in Eqs.(\ref{linFk})-(\ref{linH}).
After some calculations, one finds the relations 
\begin{eqnarray}
2 {\cal F}^{(K)} &=& h_1 + h_2 - h^{\theta_i}_{~\theta_i} 
            - { 16 \over 3} \delta \lambda,
\label{solFk} \\
2 {\cal F} &=& h_1 + h_2 - h^{\theta_i}_{~\theta_i} 
            + { 8 \over 3} \delta \lambda - 8 \delta \nu,
\label{solF} \\
2 {\cal H} &=& h_1 + h_2 - h^{\theta_i}_{~\theta_i} 
            + { 8 \over 3} \delta \lambda + 8 \delta \nu,
\label{solH}
\end{eqnarray}
where $h^{\theta_i}_{~\theta_i} = h^\chi_{~\chi} +  h^\theta_{~\theta} +
 h^\phi_{~\phi}$.
Using (\ref{solFk})-(\ref{solH}), one obtains the linearized equation
for $\delta \nu$ and $\delta \lambda$ as
\begin{eqnarray}
&&{\bar \nabla}^2 \delta \nu -{8 Q_1^2 \over r^6 f_1^2 f^{1/3}} \delta \nu =0, 
\label{nu-decoupled} \\
&&{\bar \nabla}^2 \delta \lambda  
- { d \over f^{1/3} } h^{rr} \partial^2_r \bar \lambda
+ { d \over f^{1/3}} h^{\mu\nu} \Gamma^r_{\mu\nu} \partial_r \bar \lambda
+ {2 \over r^6 f^{1/3}} \left [ {Q_1^2 \over f_1^2} 
    - {Q_K^2 \over f_K^2} \right ] h^{\theta_i}_{~\theta_i} 
\nonumber \\
&&~~~~~~
- {8 \over 3 r^6 f^{1/3}} \left [ {Q_1^2 \over f_1^2} 
    + 2 {Q_K^2 \over f_K^2} \right ] \delta \lambda =0.
\label{lambda-decoupled} 
\end{eqnarray}
We wish to point out that $\delta \nu$-equation is decoupled completely but 
$\delta \lambda$-equation still remains a coupled form.

\subsection{Dilaton Gauge}
We recognize that it is not enough to decouple $\delta \lambda$-
equation from the harmonic gauge condition. But if one introduces the 
dilaton gauge(
${\bar \nabla}_\mu {\hat h}^{\mu \rho}=h^{\mu\nu}\Gamma^\rho_{\mu\nu}$) 
\cite{Lee98-1}, 
the $\delta \lambda$-equation can be 
reduced to a better simple form.  Under this gauge, one finds 
the same relations as those in Eqs.(\ref{solFk})-(\ref{solH}) and 
the same equation for $\delta \nu$ as in Eq.(\ref{nu-decoupled}).
One finds the $\delta \lambda$-equation
\begin{eqnarray}
&&{\bar \nabla}^2 \delta \lambda  
- { d \over f^{1/3} } h^{rr} \partial^2_r \bar \lambda
+ {2 \over r^6 f^{1/3}} \left [ {Q_1^2 \over f_1^2} 
    - {Q_K^2 \over f_K^2} \right ] h^{\theta_i}_{~\theta_i} 
- {8 \over 3 r^6 f^{1/3}} \left [ {Q_1^2 \over f_1^2} 
    + 2 {Q_K^2 \over f_K^2} \right ] \delta \lambda  =0.
\label{lambda-decoupled1} 
\end{eqnarray}
In order to decouple the second term($h^{rr}$) in 
Eq.(\ref{lambda-decoupled1}), one use the Einstein's equation.
However, it seems to be a non-trivial task.  This is because 
($t,r$)-component of Eq.(\ref{linR}) gives rise to the 
second order differential equation for $h_2$.  
Instead, one may choose $h_2=h^{\theta_i}_{~\theta_i} =0$, 
which is compatible with the dilaton gauge.  Then, 
in the dilute gas limit, we find a new equation for $\delta \lambda$,
\begin{equation}
\bar \nabla ^2 \delta \lambda - 
    { 8 Q_1^2 \over 3 r^6 f_1^2 f^{1/3} } \delta \lambda =0,
\label{lambda-dilute}
\end{equation}
which is similar to Eq.(\ref{nu-decoupled}).
The situation may be getting better when one introduces the simplest 
choice such as the K-K setting\cite{Kra97}.

\subsection{Krasnitz-Klebanov Setting}
In this case, the metric perturbation $h_{\mu\nu}$ takes the form
\begin{equation}
h^\mu_{~\nu} = {\rm diag} \left [ 
       h_1, h_2, 0,0,0 \right ].
\label{ptr-KK}
\end{equation}
Under this setting, the harmonic gauge condition leads to
\begin{equation}
{1 \over 2} ( h_1 -h_2)' = 
  \left ( {1 \over 2} { d' \over d} + {3 \over r} 
  + {1 \over 6} { f' \over f} \right ) h_2 
 - \left ( { 1\over 2} { d' \over d} - {1 \over 3} { f' \over f} 
       \right ) h_1,
\label{harmonic-con}
\end{equation}
where the prime($'$) means the differentiation with respect to $r$. 
On the other hand, the dilaton gauge condition gives us the relation,
\begin{equation}
{1 \over 2} ( h_1 -h_2)' = 
  \left (  { d' \over d} + {3 \over r} \right ) h_2. 
\label{dilaton-con}
\end{equation}

From now on our calculation will be performed without any 
gauge choice for $h_{\mu\nu}$ and restriction on charges.
Solving Eqs.(\ref{linFk})-(\ref{linH}), one can express 
three U(1) fields in terms of $\delta \lambda , \delta \nu, 
h_1, h_2$ as
\begin{eqnarray}
&2{\cal F}^{(K)} &= h_1 + h_2  - {16 \over 3 } \delta \lambda, 
\label{sol-fk} \\
&2{\cal F} &= h_1 + h_2  + { 8 \over 3} \delta \lambda - 8 \delta \nu, 
\label{sol-f} \\
&2{\cal H} &= h_1 + h_2  + {8 \over 3} \delta \lambda + 8 \delta \nu.
\label{sol-h}
\end{eqnarray}
These are consistent with Eqs.(\ref{solFk})-(\ref{solH}) when 
$h^{\theta_i}_{~\theta_i} = 0$.
Ten off-diagonal elements of Einstein equation (\ref{linR})
are given by
\begin{eqnarray}
(t,r) : & {1 \over 4} \left( 
{f' \over f} +  {6 \over r} \right ) 
\partial_t h_2 + {1 \over 3}
\left ({ f_1' \over f_1} +{f_5' \over f_5} - 2 {f_K' \over f_K} \right ) \partial_t \delta \lambda 
- \left ( {f_1' \over f_1} - { f_5' \over f_5} \right ) \partial_t \delta \nu = 0, 
\label{eqtr}\\
(t,\chi) : & \partial_t \partial_\chi h_2 = 0, \label{eqtchi}\\
(t,\theta) : & \partial_t \partial_\theta h_2 = 0, \label{eqtth}\\
(t,\phi) :   & \partial_t \partial_\phi h_2 = 0, \label{eqtf}\\
(r,\chi) : & -{ 1 \over 2}( \partial_r - 3 \Gamma^\chi_{r\chi}) 
\partial_\chi h_1  - \left ( {1 \over 4} { d' \over d} + { 1 \over r} \right ) \partial_\chi (h_1 - h_2 ) \nonumber \\ &+
{ 1 \over 3} \left( {f_1' \over f_1} + {f_5' \over f_5} - 
2 {f_K' \over f_K} \right ) 
\partial_\chi \delta \lambda -
\left ( { f_1' \over f_1} - {f_5' \over f_5} \right ) 
\partial_\chi \delta \nu = 0,  \label{eqrchi}\\ 
(r,\theta) : & -{ 1 \over 2}( \partial_r - 3 \Gamma^\theta_{r\theta}) 
\partial_\theta h_1  - \left ( {1 \over 4} { d' \over d} + { 1 \over r} \right ) \partial_\theta (h_1 - h_2 ) \nonumber \\ &+
{ 1 \over 3} \left( {f_1' \over f_1} + {f_5' \over f_5} - 
2 {f_K' \over f_K} \right ) 
\partial_\theta \delta \lambda -
\left ( { f_1' \over f_1} - {f_5' \over f_5} \right ) 
\partial_\theta \delta \nu = 0,  \label{eqrth}\\ 
(r,\phi) : & -{ 1 \over 2}( \partial_r - 3 \Gamma^\phi_{r\phi}) 
\partial_\phi h_1  - \left ( {1 \over 4} { d' \over d} + { 1 \over r} \right ) \partial_\phi (h_1 - h_2 ) \nonumber \\ &+
{ 1 \over 3} \left( {f_1' \over f_1} + {f_5' \over f_5} - 
2 {f_K' \over f_K} \right ) 
\partial_\phi \delta \lambda -
\left ( { f_1' \over f_1} - {f_5' \over f_5} \right ) 
\partial_\phi \delta \nu = 0,  \label{eqrf}\\ 
(\chi,\theta) : &  ( \partial_\chi - 
\Gamma^\theta_{\chi\theta}) \partial_\theta( h_1 + h_2)= 0, \label{eqchith}\\
(\chi,\phi) : &  ( \partial_\chi - 
\Gamma^\phi_{\chi\phi}) \partial_\phi( h_1 + h_2)= 0, \label{eqchif} \\
(\theta,\phi) : &  ( \partial_\theta - 
\Gamma^\phi_{\theta\phi}) \partial_\phi( h_1 + h_2)= 0. \label{eqtthf}
\end{eqnarray}
And five diagonal elements of (\ref{linR}) take the form
\begin{eqnarray}
&(t,t) :& -{1 \over 2} { f \over d^2}\partial^2_t h_2 + 
{ 1 \over 2} \partial^2_r h_1 + 
{3 \over 2}{1 \over r} h_1' \nonumber \\ 
&&+ {1 \over 2} { 1 \over d} { 1 \over r^2} \left [ \partial^2_\chi + 
2 \cot \chi \partial_\chi +
{1 \over \sin^2 \chi}( \partial^2_\theta + \cot \theta \partial_\theta + 
{1 \over \sin^2 \theta} \partial^2_\phi ) \right ] h_1 \nonumber \\
&&+{ 1 \over 6}\left({ f_1' \over f_1} +
{f_5' \over f_5} +{f_K' \over f_K} \right) (h_2'- h_1') 
+ {1 \over 3} { d' \over d} \left({ f_1' \over f_1} +
{f_5' \over f_5} +{f_K' \over f_K} \right )(h_2 - h_1) \nonumber \\ 
&&-{ 1 \over 3}\left({ f_1'^2 \over f_1^2} +
{f_5'^2 \over f_5^2} +{f_K'^2 \over f_K^2} \right) (h_2-h_1) 
- {1 \over 4} { d' \over d} (h_2' - 3 h_1') \nonumber \\ 
&& +{4 \over 3}{ Q_K^2 \over r^6 f_K^2}{1 \over d}( h_2 - 2  {\cal F}^{(K)}
        - {8\over 3} \delta \lambda) 
 +{4 \over 3}{ Q_1^2 \over r^6 f_1^2}{1 \over d}( h_2 - 2  {\cal F}
        + {4\over 3} \delta \lambda-4 \delta \nu) \nonumber \\
&& +{4 \over 3}{ Q_5^2 \over r^6 f_5^2}{1 \over d}( h_2 - 2  {\cal H}
        + {4 \over 3} \delta \lambda+4 \delta \nu) =0,
\label{eqtt} 
\\ 
&(r,r) :& -{1 \over 2} { f \over d^2}\partial^2_t h_2 + 
{ 1 \over 2} \partial^2_r h_1 -  
{3 \over 2}{1 \over r} h_2' \nonumber \\ 
&&+ {1 \over 2} { 1 \over d} { 1 \over r^2} \left [ \partial^2_\chi + 
2 \cot \chi \partial_\chi +
{1 \over \sin^2 \chi}( \partial^2_\theta + \cot \theta \partial_\theta + 
{1 \over \sin^2 \theta} \partial^2_\phi ) \right ] h_2 \nonumber \\
&&-{ 1 \over 12}\left({ f_1' \over f_1} +
{f_5' \over f_5} +{f_K' \over f_K} \right) (5 h_1'+h_2') 
- {2 \over 3} \left({ f_1' \over f_1} +
{f_5' \over f_5} - 2 {f_K' \over f_K} \right ) 
\delta \lambda' \nonumber \\ 
&&+ 2 \left({ f_1' \over f_1} -
{f_5' \over f_5} \right) \delta \nu' 
- {1 \over 4} { d' \over d} (h_2' - 3 h_1') 
 +{4 \over 3}{ Q_K^2 \over r^6 f_K^2}{1 \over d}( h_1 - 2  {\cal F}^{(K)}
        - {8\over 3} \delta \lambda) \nonumber \\
&& +{4 \over 3}{ Q_1^2 \over r^6 f_1^2}{1 \over d}( h_1 - 2  {\cal F}
        + {4\over 3} \delta \lambda-4 \delta \nu) 
 +{4 \over 3}{ Q_5^2 \over r^6 f_5^2}{1 \over d}( h_1 - 2  {\cal H}
        + {4 \over 3} \delta \lambda+4 \delta \nu) =0,
\label{eqrr}  
\\
&(\chi,\chi) :&{1 \over r^2}{1 \over d} \partial^2_\chi ( h_1 + h_2) \nonumber \\ 
&&- { 2 \over r} \left ( {2\over r}+ { d' \over d} \right ) h_2 
- {1 \over 3} { d' \over d} { f' \over f} h_2
- {1 \over 6} \left ( {f' \over f} + { 6 \over r}\right ) (h_2' - h_1')
+ { 1 \over 3} \left( {f_1'^2 \over f_1^2} + {f_5'^2 \over f_5^2}
   + { f_K'^2 \over f_K^2} \right) h_2 
\nonumber \\
&& -{ 4 \over 3} { Q_K^2 \over r^6 f_K^2}{ 1\over d}
( h_1 + h_2 -2 {\cal F}^{(K)} - { 8 \over 3} \delta \lambda)
 -{ 4 \over 3} { Q_1^2 \over r^6 f_1^2}{ 1\over d}
( h_1 + h_2 -2 {\cal F} + { 4 \over 3} \delta \lambda-4 \delta \nu)\nonumber \\
&& -{ 4 \over 3} { Q_5^2 \over r^6 f_5^2}{ 1\over d}
( h_1 + h_2 -2 {\cal H} + { 4 \over 3} \delta \lambda+4 \delta \nu)=0,
\label{eqchichi}  
\\
&(\theta,\theta) :&{1 \over r^2 \sin^2 \chi}{1 \over d} \partial^2_\theta ( h_1 + h_2) 
+{ 1\over {r^2d}} \cot \chi \partial_\chi (h_1+h_2) \nonumber \\
&&- { 2 \over r} \left ( {2\over r}+ { d' \over d} \right ) h_2 
- {1 \over 3} { d' \over d} { f' \over f} h_2
- {1 \over 6} \left ( {f' \over f} + {6 \over r}\right ) (h_2' - h_1')
+ { 1 \over 3} \left( {f_1'^2 \over f_1^2} + {f_5'^2 \over f_5^2}
   + { f_K'^2 \over f_K^2} \right) h_2 
\nonumber \\
&& -{ 4 \over 3} { Q_K^2 \over r^6 f_K^2}{ 1\over d}
( h_1 + h_2 -2 {\cal F}^{(K)} - { 8 \over 3} \delta \lambda)
 -{ 4 \over 3} { Q_1^2 \over r^6 f_1^2}{ 1\over d}
( h_1 + h_2 -2 {\cal F} + { 4 \over 3} \delta \lambda-4 \delta \nu)\nonumber \\
&& -{ 4 \over 3} { Q_5^2 \over r^6 f_5^2}{ 1\over d}
( h_1 + h_2 -2 {\cal H} + { 4 \over 3} \delta \lambda+4 \delta \nu)=0,
\label{eqthth}  
\\
&(\phi,\phi) :&{1 \over r^2 \sin^2 \chi \sin^2 \theta} 
{ 1 \over d}\partial^2_\phi ( h_1 + h_2) 
+ {1 \over {r^2d}}\cot \chi \partial_\chi (h_1+h_2)
+ {1 \over {r^2d}}{1 \over \sin^2 \chi} \cot\theta \partial_\theta (h_1 + h_2) \nonumber \\
&&- { 2 \over r} \left ( {2\over r}+ { d' \over d} \right ) h_2 
- {1 \over 3} { d' \over d} { f' \over f} h_2  
- {1 \over 6} \left ( {f' \over f} + {6 \over r}\right ) (h_2' - h_1')
+ { 1 \over 3} \left( {f_1'^2 \over f_1^2} + {f_5'^2 \over f_5^2}
   + { f_K'^2 \over f_K^2} \right) h_2 
\nonumber \\
&& -{ 4 \over 3} { Q_K^2 \over r^6 f_K^2}{ 1\over d}
( h_1 + h_2 -2 {\cal F}^{(K)} - { 8 \over 3} \delta \lambda)
 -{ 4 \over 3} { Q_1^2 \over r^6 f_1^2}{ 1\over d}
( h_1 + h_2 -2 {\cal F} + { 4 \over 3} \delta \lambda-4 \delta \nu)\nonumber \\
&& -{ 4 \over 3} { Q_5^2 \over r^6 f_5^2}{ 1\over d}
( h_1 + h_2 -2 {\cal H} + { 4 \over 3} \delta \lambda+4 \delta \nu)=0,
\label{eqff} 
\end{eqnarray}
The fixed scalar equations (\ref{lin-nu}) and (\ref{lin-lambda}) lead to
\begin{eqnarray}
&&-{f \over d^2} \partial^2_t \delta \nu + \left [ \partial_r^2 +
\left( { d' \over d} + {3 \over r} \right ) \partial_r \right ] \delta \nu
\nonumber \\
&&~~+ { 1 \over d}{1 \over r^2} \left [ \partial^2_\chi + 
2 \cot\chi \partial_\chi + {1 \over \sin^2\chi} 
\left ( \partial^2_\theta + \cot \theta \partial_\theta +
{1 \over \sin^2\theta} \partial^2_\phi \right )
\right ] \delta \nu \nonumber \\
&&~~-{1 \over 4} { d' \over d} \left({ f_1' \over f_1} 
-{f_5' \over f_5}\right) h_2  
-{ 1 \over 8} \left({ f_1' \over f_1} 
-{f_5' \over f_5}\right) (h_2'-h_1') 
+ { 1 \over 4} \left( {f_1'^2 \over f_1^2} - {f_5'^2 \over f_5^2} 
\right )h_2 \nonumber \\
&&~~-{ Q_1^2 \over r^6 f_1^2}{1 \over d}\left(h_1 + h_2  -2 {\cal F}
+ {4 \over 3} \delta \lambda -4 \delta \nu\right) 
+{ Q_5^2 \over r^6 f_5^2}{1 \over d}\left(h_1 + h_2  -2 {\cal H}
+ {4 \over 3} \delta \lambda +4 \delta \nu\right)=0,
\label{eqnu}
\\
&&-{f \over d^2} \partial^2_t \delta \lambda + \left [ \partial_r^2 +
\left( { d' \over d} + {3 \over r} \right ) \partial_r \right ] \delta \lambda
\nonumber \\
&&~~+ { 1 \over d}{1 \over r^2} \left [ \partial^2_\chi + 
2 \cot\chi \partial_\chi + {1 \over \sin^2\chi} 
\left ( \partial^2_\theta + \cot \theta \partial_\theta +
{1 \over \sin^2\theta} \partial^2_\phi \right )
\right ] \delta \lambda \nonumber \\
&&~~+{1 \over 4} { d' \over d} \left({ f_1' \over f_1} 
+{f_5' \over f_5}-2 {f_K' \over f_K}\right) h_2  
+{ 1 \over 8} \left({ f_1' \over f_1} 
+{f_5' \over f_5}-2 {f_K' \over f_K}\right) (h_2'-h_1') 
- { 1 \over 4} \left( {f_1'^2 \over f_1^2} + {f_5'^2 \over f_5^2} 
-2 {f_K'^2 \over f_K^2}\right )h_2 \nonumber \\
&&~~-2{ Q_K^2 \over r^6 f_K^2}{1 \over d}\left(h_1 + h_2  -2 {\cal F}^{(K)}
- {8 \over 3} \delta \lambda\right) \nonumber \\
&&~~+{ Q_1^2 \over r^6 f_1^2}{1 \over d}\left(h_1 + h_2  -2 {\cal F}
+ {4 \over 3} \delta \lambda -4 \delta \nu\right) 
+{ Q_5^2 \over r^6 f_5^2}{1 \over d}\left(h_1 + h_2  -2 {\cal H}
+ {4 \over 3} \delta \lambda +4 \delta \nu\right)=0.
\label{eqlambda}
\end{eqnarray}

\section{S-wave Propagations}
\label{propagation}
From the Bianchi identities (\ref{bianchi}) one has 
\begin{eqnarray}
&&\partial_\chi {\cal F}^{(K)} = \partial_\theta {\cal F}^{(K)} =
      \partial_\phi {\cal F}^{(K)}=0, \nonumber \\
&&\partial_\chi {\cal F} = \partial_\theta {\cal F} =
      \partial_\phi {\cal F}=0, \nonumber \\
&&\partial_\chi {\cal H} = \partial_\theta {\cal H} =
      \partial_\phi {\cal H}=0. \label{eq-bianchi} 
\end{eqnarray}
This implies either ${\cal F}^{(K)}={\cal F}^{(K)} (t,r), {\cal F}=
{\cal F}(t,r), {\cal H}={\cal H}(t,r)$ or 
${\cal F}^{(K)}={\cal F}= {\cal H}=0$.  
The latter together with (\ref{sol-fk})-(\ref{sol-h}) means 
that all higher modes of $l \ge 1$ are forbidden in this scheme.  
We wish to study 
the s-wave propagation with the first case.  This case dominates 
in the absorption of low energies. 
The important 
one can be derived from ($t,r$)-component of the 
Einstein's equation.  
By integrating (\ref{eqtr}) over time, we can obtain the relation
\begin{equation}
\left ( {f' \over f} + {6 \over r} \right ) h_2 =
 -{4 \over 3} \left ( {f_1' \over f_1} + {f_5' \over f_5} -2 {f_K' \over f_K} 
\right ) \delta \lambda + 4 \left ( {f_1' \over f_1} 
  - {f_5' \over f_5} \right ) \delta \nu.
\label{relation-h2}
\end{equation}
From three angular equations 
(\ref{eqchichi})-(\ref{eqff}), one finds the relation
\begin{eqnarray}
&&\left ( {f' \over f} + {6 \over r} \right ) (h_1' - h_2' ) =
  \left [ 
2 {d' \over d} {f' \over f} 
+ { 12 \over r} \left ( { 2 \over r} + {d' \over d} \right ) 
-2 \left ( {f_1'^2 \over f_1^2} + {f_5'^2 \over f_5^2} 
+ {f_K'^2 \over f_K^2} \right )
\right ] h_2
\nonumber \\
&&~~~~~~~~~~~~~~~~~~~~~~~~~~~~~
+{32 \over 3} { 1 \over {r^6 d}} \left (
2 {Q_K^2 \over f_K^2} - {Q_1^2 \over f_1^2} - {Q_5^2 \over f_5^2} \right )
\delta \lambda
+ 32 { 1 \over {r^6 d}} \left ( {Q_1^2 \over f_1^2} 
- {Q_5^2 \over f_5^2} \right )\delta \nu.
\label{relation-h1mh2}
\end{eqnarray}
Eqs.(\ref{eqtt}) and (\ref{eqrr}) lead to another relation 
\begin{eqnarray}
&&\left ( {f' \over f} + { 6 \over r} \right ) ( h_1' + h_2') =
-{ 8 \over 3} \left ( { f_1' \over f_1} + {f_5' \over f_5} 
-2{f_K' \over f_K} \right ) \delta \lambda' 
+ 8 \left ( {f_1' \over f_1} - { f_5' \over f_5} \right )
 \delta \nu '
\label{relation-h1ph2}
\end{eqnarray}
From 
(\ref{relation-h1mh2}) and (\ref{relation-h1ph2}), one can obtain
\begin{eqnarray}
&&\left ( {f' \over f} + {6 \over r} \right ) h_2'  =
-{ 4 \over 3} \left ( { f_1' \over f_1} + {f_5' \over f_5} 
-2{f_K' \over f_K} \right ) \delta \lambda '
+ 4 \left ( {f_1' \over f_1} - { f_5' \over f_5} \right )
\delta \nu '
\nonumber \\
&&~~~~~~~~~~~~~~~~~~~~~~~~~~~
  - \left [ 
 {d' \over d} {f' \over f} 
+ { 6 \over r} \left ( { 2 \over r} + {d' \over d} \right ) 
- \left ( {f_1'^2 \over f_1^2} + {f_5'^2 \over f_5^2} 
+ {f_K'^2 \over f_K^2} \right )
\right ] h_2
\nonumber \\
&&~~~~~~~~~~~~~~~~~~~~~~~~~~~
-{16 \over 3} { 1 \over {r^6 d}} \left (
2 {Q_K^2 \over f_K^2} - {Q_1^2 \over f_1^2} - {Q_5^2 \over f_5^2} \right )
\delta \lambda
- 16 { 1 \over {r^6 d}} \left ( {Q_1^2 \over f_1^2} 
- {Q_5^2 \over f_5^2} \right )\delta \nu.
\label{relation-h2p}
\end{eqnarray}
However, this equation is redundant because it can be obtained by differentiating 
(\ref{relation-h2}) with respect to $r$.  
All information for $h_1$ and $h_2$ are thus encoded in (\ref{relation-h2}) 
and (\ref{relation-h1mh2}), which say that   
$h_1$ and $h_2$ are not the independent modes  and thus only two 
fixed scalars are propagating in the 5D black hole background.  
This can be confirmed from the fact that the relevant value 
of $l$ should be determined by 
$l \ge |S|$, $S$=spin.  Since the gravitons have spin 2, it 
is not surprising that they are redundant with 
$l=0$(s-wave) case.  Similarly, 
three U(1) modes with $l=0$ are also redundant because the photon has 
spin 1.  This was clearly shown in (\ref{sol-fk})-(\ref{sol-h}).
Inserting (\ref{relation-h2}), (\ref{relation-h1mh2}) into (\ref{eqnu}) and 
(\ref{eqlambda}), we obtain the following equations:
\begin{eqnarray}
&&\left[ r^{-3} \partial_r d r^3 \partial_r - d^{-1} f \partial^2_t 
+f_{\nu\nu}(r) \right ] \delta \nu + f_{\nu\lambda}(r) \delta \lambda =0,
\label{sub-eqnu}\\
&&\left[ r^{-3} \partial_r d r^3 \partial_r - d^{-1} f \partial^2_t 
+f_{\lambda\lambda}(r) \right ] \delta \lambda + f_{\lambda\nu}(r) \delta \nu =0,
\label{sub-eqlambda}
\end{eqnarray}
where $f_{\nu\nu}(r),~f_{\nu\lambda}(r),~f_{\lambda\lambda}(r),~
f_{\lambda\nu}(r)$ are given by
\begin{eqnarray}
&&f_{\nu\nu}(r)=-{ 8 \over {r^2[3r^4 + 2 r^2(r_1^2+r_5^2 + r_K^2)+
r_1^2 r_K^2 +r_1^2 r_5^2 + r_5^2 r_K^2]^2}} \times  \nonumber \\
&&~~~~~~~~~~~~\left [
3r^4 \left\{r_1^4 + r_5^4 + r_1^2 r_5^2 +{3 \over 2}r_0^2(r_1^2 +r_5^2)\right \}
 \right . \nonumber \\
&&~~~~~~~~~~~~~ 
+3r^2 \left \{r_5^4r_K^2 + r_1^4 r_K^2 + r_1^4 r_5^2 + r_5^4 r_1^2 
  +2 (r_1r_5r_K)^2 +2 r_0^2(r_1^2r_K^2 + r_1^2 r_5^2 + r_5^2 r_K^2 )\right \}
\nonumber \\
&&~~~~~~~~~~~~~ 
+r_1^4 r_5^4 + r_5^4 r_K^4 + r_1^4 r_K^4 +2 r_1^2 r_5^2 r_K^2 (r_1^2 +r_5^2 
+ r_K^2)  
\nonumber \\
&&~~~~~~~~~~~~~~~~~ \left .
+ r_0^2 \left \{ 3 r_1^2 r_5^2 r_K^2 + {1 \over 2} r_1^4 (r_5^2 + r_K^2) 
+ {1 \over 2} r_5^4 (r_1^2 + r_K^2) + 2 r_K^4 ( r_1^2 + r_5^2) \right \}
\right ], \label{f-nunu} \\
&&f_{\nu\lambda}(r)={ 8 \over {r^2[3r^4 + 2 r^2(r_1^2+r_5^2 + r_K^2)+
r_1^2 r_K^2 +r_1^2 r_5^2 + r_5^2 r_K^2]^2}} \times  \nonumber \\
&&~~~~~~~~~~~~\left [
r^4 \left\{r_1^4 - r_5^4 -r_5^2 r_K^2 + r_1^2 r_K^2 +
{3 \over 2}r_0^2(r_1^2 -r_5^2)\right \}
 \right . \nonumber \\
&&~~~~~~~~~~~~~\left . 
+r^2 \left \{r_1^4(r_5^2 + r_K^2) - r_1^2 r_5^4 - r_5^4 r_K^2  \right \}
+{1 \over 2} r_0^2 \left \{ r_1^2 r_5^2 (r_5^2 -r_1^2)  
+ r_K^2 (r_5^4 - r_1^4) \right \}
\right ], \label{f-nulambda} \\
&&f_{\lambda\nu}(r)=3 f_{\nu\lambda}, \label{f-lambdanu}  \\
&&f_{\lambda\lambda}(r)=-{ 8 \over {r^2[3r^4 + 2 r^2(r_1^2+r_5^2 + r_K^2)+
r_1^2 r_K^2 +r_1^2 r_5^2 + r_5^2 r_K^2]^2}} \times  \nonumber \\
&&~~~~~~~~~~~~\left [
r^4 \left\{r_1^4 + r_5^4 - r_1^2 r_5^2 +  4 r_K^4 + 2 r_K^2 (r_1^1 + r_5^2) +
{3 \over 2}r_0^2(r_1^2 +r_5^2 + 4 r_K^2)\right \}
 \right . \nonumber \\
&&~~~~~~~~~~~~~ 
+r^2 \left \{r_5^4r_K^2 + r_1^4 r_K^2 + r_1^4 r_5^2 + r_1^2 r_5^4  
+ 4 r_K^4 (r_1^2 + r_5^2)  +6 (r_1r_5r_K)^2 \right .
\nonumber \\
&&~~~~~~~~~~~~~~~~~~~~~~~~~ \left .
   +6 r_0^2(r_1^2r_K^2 + r_1^2 r_5^2 + r_5^2 r_K^2 ) \right \}
\nonumber \\
&&~~~~~~~~~~~~~ 
+r_1^4 r_5^4 + r_5^4 r_K^4 + r_1^4 r_K^4 +2 r_1^2 r_5^2 r_K^2 (r_1^2 +r_5^2 
+ r_K^2)  
\nonumber \\
&&~~~~~~~~~~~~~~~~~ \left .
+ r_0^2 \left \{ 3 r_1^2 r_5^2 r_K^2 + {3 \over 2} r_1^4 (r_5^2 + r_K^2) 
+ {3 \over 2} r_5^4 (r_1^2 + r_K^2)  \right \}
\right ]. \label{f-lambdalambda} 
\end{eqnarray}
Note that for $r_1=r_5 \equiv R$, one finds 
$f_{\nu\lambda}=f_{\lambda\nu}=0$.  
Then equations (\ref{sub-eqnu}) and (\ref{sub-eqlambda}) 
reduce to 
\begin{eqnarray}
&&\left[ r^{-3} \partial_r d r^3 \partial_r - d^{-1} f \partial^2_t 
-{{8 R^4} \over {r^2(r^2 + R^2)^2}}\left ( 1 + {r_0^2 \over R^2} 
\right ) \right ] \delta \nu  =0,
\label{reduce-eqnu}\\
&&\left[ r^{-3} \partial_r d r^3 \partial_r - d^{-1} f \partial^2_t 
-{{8 (R^2+2 r_K^2)^2} \over {r^2(3r^2 + (R^2+ 2r_K^2))^2}}
\left ( 1 + {3r_0^2 \over {R^2+2r_K^2}} 
\right ) \right ] \delta \lambda =0.
\label{reduce-eqlambda}
\end{eqnarray}
We note that Eq.(\ref{reduce-eqnu}) is exactly the same form as in 
Eq.(\ref{nu-decoupled}).  This is so because for $r_1=r_5$, 
there is no mixing between graviton
and fixed scalar($\delta \nu$).  However, a mixing between
graviton and $\delta \lambda$ is still present and thus we obtain
the decoupled equation (\ref{reduce-eqlambda}) by using 
(\ref{relation-h2}) and (\ref{relation-h1mh2}).  
We would like to find the fixed 
scalar equations for the general ($r_1 \neq r_5 \neq r_K$) case.  
Eqs.(\ref{sub-eqnu}) and (\ref{sub-eqlambda}) 
can be modified with $3f_{\nu\lambda}(r)=f_{\lambda\nu}(r)$ and 
$\delta \tilde \lambda\equiv{\delta \lambda / \sqrt{3}}$ as
\begin{eqnarray}
&&\left[ r^{-3} \partial_r d r^3 \partial_r - d^{-1} f \partial^2_t 
+f_{\nu\nu}(r) \right ] \delta \nu + 
\sqrt{3}f_{\nu\lambda}(r) \delta \tilde \lambda =0,
\label{sub-eqnu1}\\
&&\left[ r^{-3} \partial_r d r^3 \partial_r - d^{-1} f \partial^2_t 
+f_{\lambda\lambda}(r) \right ] \delta \tilde \lambda + 
\sqrt{3} f_{\nu\lambda}(r) \delta \nu =0.
\label{sub-eqlambda1}
\end{eqnarray}
The above can be decoupled by a rotation of the fields as
\begin{eqnarray}
&& \delta \tilde \lambda = (\cos \alpha) \phi_+ + (\sin \alpha) \phi_-,\\
&& \delta \nu = -(\sin \alpha) \phi_+ + (\cos \alpha) \phi_-,
\end{eqnarray}
where the rotation angle ($\alpha$) satisfies the relation
\begin{equation}
\tan \alpha - {1 \over \tan \alpha} = { 1 \over {\sqrt{3}}} 
{{f_{\lambda\lambda}(r)-f_{\nu\nu}(r)} \over f_{\nu\lambda}(r)}
={2 \over \sqrt{3}}{{r_1^2 + r_5^2 -2 r_K^2} 
\over {r_1^2 -r_5^2}}.
\label{tangent}
\end{equation}
From (\ref{tangent}) one obtains
\begin{equation}
\cos^2 \alpha = { 1\over 2} \pm { 1 \over 4} { r_1^2 + r_5^2 -2 r_K^2 \over
   \sqrt{r_1^4 + r_5^4 + r_K^4 -r_1^2 r_5^2 - r_1^2 r_K^2 -r_5^2 r_K^2} }.
\label{cosine}
\end{equation}
Then (\ref{sub-eqnu1}) and (\ref{sub-eqlambda1}) lead to the decoupled 
equations for $\phi_\pm$,
\begin{eqnarray}
&&\left [ r^{-3} \partial_r d r^3 \partial_r - d^{-1}f \partial_t^2 +
  \sin^2 \alpha f_{\nu\nu} + \cos^2 \alpha f_{\lambda\lambda} -
  2 \sqrt{3} \cos \alpha \sin \alpha f_{\nu \lambda} \right ] \phi_+=0,
\label{eq-plus} \\
&&\left [ r^{-3} \partial_r d r^3 \partial_r - d^{-1}f \partial_t^2 +
  \cos^2 \alpha f_{\nu\nu} + \sin^2 \alpha f_{\lambda\lambda} +
  2 \sqrt{3} \cos \alpha \sin \alpha f_{\nu \lambda} \right ] \phi_-=0.
\label{eq-minus}
\end{eqnarray}
Here we consider $\phi_\pm(r,t) = \tilde \phi_\pm(r) e^{-i \omega t}$ 
as a mode with energy $\omega$.  
Inserting (\ref{cosine}), (\ref{f-nunu})-(\ref{f-lambdalambda}) into
(\ref{eq-plus})-(\ref{eq-minus}), we obtain the equations
\begin{equation}
\left [ ( d r^3 \partial_r)^2 + \omega^2 r^6 f - 
  {{8 d r^4 r_{\pm}^4} \over (r^2 + r_{\pm}^2)^2 }
  \left ( 1 + {r_0^2 \over r_{\pm}^2} \right ) \right ] \tilde \phi_{\pm}=0,
\label{eq-tot} 
\end{equation}
where the effective radii $r_{\pm}$ are defined as
\begin{eqnarray}
&&r_\pm^2 = {1 \over 3} \left [ r_1^2 + r_5^2 + r_K^2 \pm 
\sqrt{r_1^4+r_5^4+r_K^4 -r_1^2r_5^2 -r_1^2 r_K^2 -r_5^2 r_K^2}\right ].
\label{r-plus} 
\end{eqnarray}
Eq.(\ref{eq-tot}) takes the 
same form as in Eq.(\ref{reduce-eqnu}).  
Since it is difficult to find an analytic solution to (\ref{eq-tot}),
we patch together a solution between the near region 
(region I, $r \ll r_1,r_5$), 
the intermediate region (region II, $r_0 \ll r \ll \omega^{-1}$) and the 
far region (region III, $r \gg r_1,r_5$).  
The region II overlaps 
each of other two because of $r_0 \ll r_1, r_5 \ll \omega^{-1}$.  
In the dilute gas regime ($r_0, r_K \ll r_1, r_5$), we write down 
the dominant terms and their approximate solutions in the three regions as
\begin{eqnarray}
&{\rm I}.&\left [ (d r^3 \partial_r)^2 + r_1^2 r_5^2 (r^2 + r_K^2) \omega^2 
      -8 r^4 d \right ]\tilde \phi_\pm^I =0, ~~~~
        \tilde \phi_\pm^I = E { r^2 \over r_0^2} +G;
\label {eq-regionI} \\
&{\rm II}.&\left [  ( r^3 \partial_r)^2  
-8 {r_\pm^4 \over \left (1 + {r_\pm^2 \over r^2}\right )^2} \right ]
\tilde \phi_\pm^{II} =0, ~~~~
  \tilde \phi_\pm^{II} ={C_\pm \over \left (1 + {r_\pm^2 \over r^2}\right )} +
      D_\pm \left ( 1+ {r_\pm^2 \over r^2}\right )^2;
\label {eq-regionII} \\
&{\rm III}.&\left [ ( r^3 \partial_r)^2 + r^6 \omega^2
 \right ] \tilde \phi_\pm^{III} =0, ~~~~
  \tilde \phi_\pm^{III} =\alpha_\pm {J_1(\omega r) \over \omega r} +
      \beta_\pm {N_1(\omega r) \over \omega r},
\label {eq-regionIII} 
\end{eqnarray}
where $C_\pm, D_\pm, \alpha_\pm, \beta_\pm$ are the unknown constants.  
The full solution in the region I can be expressed in terms of 
the hypergeometric functions \cite{Cal}, 
and we present here the limiting form for $r \gg r_0$.  $E$ 
is obtained by the requirement that the solution 
be purely ingoing at the horizon as
\begin{equation}
E = {{2 \Gamma(1 -i a - i b ) } \over {\Gamma(2 -i a) \Gamma(2 -i b) }}.
\label{coef-E}
\end{equation}
Here $a$ and $b$ are related to the 
left and right moving temperatures as
\begin{equation}
a = { \omega \over 4 \pi T_L}, ~~~~~~ b = {\omega \over 4 \pi T_R}.
\label{def-ab}
\end{equation}
The quantity $G$ may be similarly fixed, but its value is not relevant to us.  
A matching procedure leads to the relation
\begin{equation}
\alpha_\pm = 2 C_\pm = 2 E {r_\pm^2 \over r_0^2}.
\label{def-alpha}
\end{equation}
The absorption probability 
is given by the ratio of the incoming fluxes at the horizon($r=r_0$) 
and at spatial infinity($r=\infty$)\cite{Mal}.   
The flux per unit solid angle for a field $f$ is given by
\begin{equation}
F = {1 \over 2 i} ( f^* d r^3 \partial_r f - c.c).
\label{flux}
\end{equation}
The absorption probability of $\phi_\pm$ is given by
\begin{equation}
P^{\phi_\pm}_{abs} = {F_{r_0} \over F_\infty} = 2 \pi r_1r_5 
\sqrt{r_0^2 + r_K^2}\, \omega^3 {r_0^4 \over {4 |E|^2 r_\pm^4}}.
\label{prob}
\end{equation}
Then the absorption cross section is given by
\begin{equation}
\sigma^{\phi_\pm}_{abs} = { 4 \pi \over \omega^3} P^{\phi_\pm}_{abs} 
      = { { \pi^3 r_1^6 r_5^6} \over { 64 r_\pm^4 
}} \omega ( \omega^2 
+ 16 \pi^2 T^2_L)(\omega^2 + 16 \pi^2 T^2_R ) 
{ { e^{\omega \over T_H} -1 } \over {(e^{\omega \over T_L}-1)
(e^{\omega \over T_R} -1) }},
\label{sigma}
\end{equation}
which is the same form as in Ref. \cite{Kra97}.  
When $r_1=r_5= r_+ = R$, 
one finds the absorption cross section for $\nu$.  
For $r_1=r_5=R, r_-^2=R^2/3$, one gets the cross section 
for $\lambda$.

\section{Discussions}
\label{discussion}
Let us first discuss the role of a fixed scalar $\nu$.  Although $\nu$ is 
related to the scale of $T^4$, it turns out to be the 10D 
dilaton($\phi_{10}$) when $\phi_6 = \phi_{10} - 2 \nu =0$.  
For $Q_1=Q_5$ case, one finds the same linearized equation 
for the harmonic, dilaton gauge, and K-K setting.  This means that the 
fixed scalar($\nu$) gives us a gauge-invariant result.  
In the low energy limit ($\omega \to 0$), 
the s-wave absorption cross section takes the form
\begin{equation}
\sigma^\nu_{abs} = C {\cal A}_H \left ( r_0 \over R \right ) ^4,
\label{abs-nu}
\end{equation}
where $C=1/4$ for the semiclassical approach from Eq.(\ref{sigma}), 
$1/16$ for the effective string method\cite{Cal}, 
$1/12$ for the $AdS_3$-calculation\cite{Lee98}, 
and $1/4$ for the boundary CFT-calculation
\cite{Teo98}.  This means that all calculation methods lead 
to the same result, upto the numerical factors.  
In the dilute gas limit($R\ll r_0$), one finds $\sigma^\nu_{abs} \to 0$,  
whereas $\sigma^{\Phi}_{abs} = {\cal A}_H$ for a minimally 
decoupled scalar $\Phi$.  

On the other hand, $\lambda(=\nu_5 - \phi_6/2$) is entirely determined by  
the scale($\nu_5$) of the KK circle($S^1$) when $\phi_6$ is 
turned off.  The semiclassical result (\ref{sigma}) with 
$r_-^2 = R^2/3$ takes the form
\begin{equation}
\sigma^\lambda_{abs} = { 9 \over 4} {\cal A}_H \left ( {r_0 \over R} \right )^4.
\label{abs-lambda}
\end{equation}
On the effective string side, the $\lambda$-coupling is \cite{Kra97}
\begin{equation}
-{ T_{eff} \over 8} \lambda \left [ 
\partial_+ X \partial_- X \left \{ (\partial_+X)^2 
  + (\partial_-X)^2 \right \} + (\partial_+X)^2 (\partial_-X)^2 
\right ]
\label{lambda-coupling}
\end{equation}
plus the fermionic terms.  Here $T_{eff}(=1/2 \pi^2 R^2)$ is the 
effective string tension.  The last term is an operator of 
dimension (2,2) which also couples to $\nu$-fixed scalar.  
This gives  
$\sigma^\lambda_{abs} ={{\cal A}_H \over 16}\left ( {r_0 \over R} \right )^4$.  
Also there are additional contributions to the cross section 
which arise from the first two terms.  They have dimensions (3,1) and (1,3).  
The presence of these gives rise to some disagreement between the 
semiclassical and effective string cross sections even for 
$Q_1=Q_5$.  

On the semiclassical calculation, this discrepancy originates 
from a complicated mixing between $\lambda$ and other fields.  
Hence it may depend on the decoupling procedure.  
In this work we find out that $\lambda$ depends on the 
gauge choice.  For example, one obtains Eq.(\ref{lambda-decoupled}) for 
the harmonic gauge, Eq.(\ref{lambda-dilute}) for the dilaton gauge 
together with $h_2=h^{\theta_i}_{~\theta_i}=0$, and Eq.(\ref{reduce-eqlambda}) 
for K-K setting.  Furthermore, 
substituting  Eq.(\ref{relation-h2}) into Eq.(\ref{relation-h1mh2}) 
leads to 
\begin{equation}
{1 \over 2} (h_1-h_2)' =
\left [ { d' \over d} 
+ { {12 /r^2 - (2 {f'}_1^2 /f_1^2 + {f'}_K^2 / f_K^2 )} \over 
           { f'/f + 6/r}}
-{4 \over r^6 d} 
  {{Q_K^2/f_K^2 - Q_5^2 / f_1^2 } \over 
     {f'_1 / f_1 - f'_K /f_K}}
\right ] h_2.
\label{KK-con}
\end{equation}
This is a result purely from the Einstein's equation.  However it is shown that 
(\ref{KK-con}) is not compatible with either the harmonic gauge 
condition Eq.(\ref{harmonic-con}) or the dilaton gauge condition 
Eq.(\ref{dilaton-con}).  
Although the K-K setting is a convenient choice for obtaining 
the decoupled equations, it does not always guarantee the 
consistent solution.  

In conclusion, the fixed scalar $\nu$ is clearly understood as a good test 
field.  However, the role of $\lambda$ as a test field is obscure because 
it is a gauge-dependent field and gives rise to some 
disagreement for the cross section.

\section*{Acknowledgement}
This work was supported in part by the Basic Science Research Institute 
Program, Ministry of Education, Project No. BSRI-98-2413 and by Korea 
Science and Engineering Foundation (94-1400-04-02-3).


\begin{references}
\bibitem{Vaf} A. Strominger and C. Vafa, Phys. Lett. {\bf B379}, 99 (1996), 
hep-th/9601029;
              C. Callan and J. Maldacena, Nucl. Phys. {\bf B472}, 591 (1996), 
hep-th/9602043;
              G. Horowitz and A. Strominger, Phys. Rev. Lett. {\bf 77}, 
2368 (1996), hep-th/9602051.
\bibitem{Mal} J. Maldacena and A. Strominger, Phys. Rev. 
{\bf D55}, 861 (1997), hep-th/9609026; 
A. Dhar, G. Mandal, and S. Wadia, Phys. Lett. {\bf B388}, 51 (1996)
, hep-th/9605234; 
J. M. Maldacena, hep-th/9705078.
\bibitem{Das} S. Das, G. Gibbons, and S. Mathur, Phys. Rev. Lett. {\bf 78}, 
417 (1997).
\bibitem{Cal} C. Callan, S. Gubser, I. Klebanov, and A. Tseytlin, Nucl. 
Phys. {\bf B489}, 65 (1997),  het-th/9610172; 
I.R. Klebanov and M. Krasnitz, Phys. Rev. {\bf D55}, R3250 (1997).
\bibitem{Kra97} M. Krasnitz and I. Klebanov, hep-th/9703216; 
M. Taylor-Robinson, hep-th/9704172.
\bibitem{Das2} S. Das, A. Dasgupta and T. Sarkar, hep-th/9702075;
               F. Dowker, D. Kastor, and J. Traschen, hep-th/9702109;
               I. Klebanov, A. Rajaraman, and A. Tseytlin, hep-th/9704112.
\bibitem{Emp} R. Emparan, hep-th/9704204; hep-th/9706204.
\bibitem{Kal}  R. Kallosh, A. Linde, T. Ortin, A, Peet, and A. van Proeyen,
              Phys. Rev. {\bf D46}, 5278 (1992).
\bibitem{Kol} B. Kol and A. Rajaraman, Phys. Rev. {\bf D56}, 986(1997), 
hep-th/9608126; H.W. Lee, Y.S. Myung and Jin Young Kim, 
Phys. Lett. {\bf B410}, 6(1997), hep-th/9704199.
\bibitem{Mal1} J. Maldacena and A. Strominger, hep-th/9702015;
              S.S. Gubser, I. Klebanov and A.A. Tseytlin, hep-th/9703040;
              S.D. Mathur, hep-th/9704156; S.S. Gubser, hep-th/9704195.
\bibitem{Hol}  C. Holzhey and F. Wilczek, Nucl. Phys. {\bf B380}, 447 (1992).
\bibitem{Cha83} 
S. Chandrasekhar,  {\it The Mathematical Theory of Black Hole}
                    (Oxford Univ. Press, New York, 1983).
\bibitem{Mtw} C. Misner, K. Thorne, and J. Wheeler, {\it Gravitation} 
(Freeman, San Francisco, 1973).
\bibitem{Gub} S.S. Gubser, hep-th/9706100.
\bibitem{Lee} H. W. Lee , Y. S. Myung and J. Y. Kim, Class. 
Quantum Grav. {\bf 14}, 759 (1997), hep-th/9510127;
               H. W. Lee , Y. S. Myung, J. Y. Kim and D. K. Park, 
Class. Quantum Grav. {\bf 14}, L53 (1997), hep-th/9603044.
\bibitem{Gre94}
  R. Gregory and R. Laflamme, Nucl. Phys. {\bf B428}, 399 (1994).
\bibitem{Lee98-1}
  H.W. Lee, N.J. Kim, Y.S. Myung, and J.Y. Kim 
       Phys. Rev. {\bf D57}, 7361 (1998), hep-th/9801152.
\bibitem{Lee98}
  H.W. Lee, N.J. Kim, Y.S. Myung, hep-th/9805050.
\bibitem{Teo98}
  E. Teo, hep-th/9805014.
\end{references}
\end{document}